\documentclass{jaa}
\usepackage[utf8]{inputenc}
\usepackage{natbib}
\usepackage{color}
\usepackage{longtable}
\usepackage{multirow}
\usepackage{graphicx}
\usepackage{hyperref}
\newcommand\Tstrut{\rule{0pt}{2.6ex}}         
\newcommand\Bstrut{\rule[-0.9ex]{0pt}{0pt}}   

\begin{document}
\title{Infrared Polarisation Study of Lynds 1340: a case of RNO 8}
\author{Archita Rai\textsuperscript{1,2}, Shashikiran Ganesh\textsuperscript{1}}
\affilOne{\textsuperscript{1}Physical Research Laboratory, Ahmedabad, Gujarat 380009, India.\\}
\affilTwo{\textsuperscript{2}Indian Institute of Astrophysics, II Block Koramangala, Bengaluru 560034, India.}


\twocolumn[{

\maketitle

\corres{rai.archita@gmail.com, archita.rai@iiap.res.in}




\begin{abstract}
This paper describes the polarisation study of a Lynds cloud, LDN 1340, $\alpha$ = 2h32m \& $\delta$ = $73^{\circ} 00^\prime$ corresponding to galactic coordinates of $\ell=$ 130$^{\circ}$.07 $b=$ 11$^{\circ}$.6, with emphasis on the RNO 8 area. The cloud has been observed using the 1.2 m telescope at Mt.Abu Infrared Observatory, in the infrared wavelength band using the Near-Infrared Camera, Spectrograph \& Polarimeter (NICSPol) instrument. The polarimetric observations were used to map the magnetic field geometry around the region.  We combined our measurements with archival data from the 2MASS and WISE surveys.  The Gaia EDR3 \& DR3 data for the same region were used for distance, proper motion, and other astrophysical information. The analysis of the data reveals areas with ordered polarisation vectors in the region of RNO 8. The position angle measurements reveal polarisation due to dichroic extinction which is consistent with the Galactic magnetic field. The magnetic field strength was calculated for the RNO 8 region using the Chandrashekhar-Fermi method and the value estimated is $\sim$ 42$\mu$G. 
\end{abstract}


}]


\doinum{12.3456/s78910-011-012-3}
\artcitid{\#\#\#\#}
\volnum{000}
\year{0000}
\pgrange{1--}
\setcounter{page}{1}
\lp{1}

\section{Introduction}
Dark nebulae, such as the ones described by \citet{lynds}, have been relatively unexplored in the optical due to the large extinction of these regions.  Lynds catalogued these dark nebulae into 6 opacity classes based upon their visual appearance in the National Geographic Palomar Observatory Sky Survey images.\\
Polarimetry is a good way to probe the magnetic field structures  \citep{davis} in dark clouds \citep{joshi1985, Jones1989,Myers,andersson}. 
Polarimetric observations of the stars in the line of sight through the dark nebulae/molecular cloud would help to probe the magnetic field structure in the cloud and along the line of sight \citep{lazarian}. For example, \citet{eswaraiah} have studied the dark globule LDN 1225 using optical polarimetry and infrared photometry to understand the extinction and magnetic field properties of the region.\\
Infrared polarimetry would be even more appropriate in regions of high extinction. Earlier work \citep[e.g][]{wilking, Kwon_2016} using infrared polarimetry  concluded that it is a valuable tool for measuring the dichroic polarization of the background stars and those embedded within the dense clouds.\\
\citet{CF}  formulated methods to estimate the magnetic field strength in the molecular clouds. The method is independent of the Zeeman effect which is difficult to detect in molecular clouds due to smaller frequency splitting in presence of weak magnetic fields. The estimation of the mean field strength was based on the knowledge of the mean gas density, the line-of-sight velocity, and the position angle dispersion (obtained from polarisation measurements). The position angle dispersion is connected to the dispersion in the orientation of the magnetic field in the plane of the sky. The estimated magnetic field strength using this method is found to be accurate when the polarisation angle fluctuations are small \citep{ostriker}.\\
\cite{Kwon_2016} discuss the observations of GGD 27 in the Lynds 291 molecular cloud complex in the constellation of Sagittarius using the SIRPOL instrument on the IRSF telescope at Sutherland, South Africa.   They study the complex structure of the magnetic field and derive the magnetic field strength based on their polarimetric observations.\\
Lynds dark nebula 1340 (hereafter referred to as L1340), a cloud of opacity class 5, located at $\alpha=$ 2h32m, $\delta=$ $73^{\circ} 00'$ ($\ell=$ 130$^{\circ}$.07,$b=$ 11$^{\circ}$.6), has been studied in a series of papers \citep[e.g.][and references therein]{Kun1994,Kun,2016ApJ...822...79K} by M. Kun and collaborators. \citet{Kun1994} have studied L1340 as a star-forming complex in Cassiopeia using optical photometry as well as radio ($^{13}$CO and C$^{18}$O) molecular emissions. Using objective prism spectroscopy they detected 13 $H\alpha$ emission line stars.  \citet{Ganesh2001} also studied this cloud using infrared photometry with the PRLNIC3 instrument on the 1.2m Mount Abu IR telescope. \citet{2016ApJ...822...79K, Kun} studied the region in the optical \& infrared using the photometric data from SDSS, WISE, and Spitzer survey to determine the young stellar populations present in the region. They conclude that it is an isolated molecular cloud of 3700 $M_{\odot}$ 
with an intermediate star formation efficiency $\sim$ $3\%$.\\
In this paper, we present an infrared polarisation study of the L1340 region using a near-infrared imaging polarimeter (H band).  Section \ref{2} briefly describes the observational procedure followed. Section \ref{3} talks about the data reduction and analysis steps for deriving infrared polarisation measurements from the observations. Section \ref{4} describes the supplementary data sets used. Section \ref{sec:results} discusses the results obtained. A concluding summary is provided in the Section \ref{6}.

\section{Observations}
\label{2}
\subsection{Telescope and instrument}

Physical Research Laboratory (PRL) operates a 1.2 m telescope at its Mount Abu Infrared Observatory (MIRO).  MIRO is located at $24^{\circ} 39^\prime 9^{\prime}{^\prime}$ (N) latitude, $72^{\circ} 46^\prime 47{^\prime}{^\prime} $ (E) longitude at 1680~m altitude.   The observatory location favours good conditions for IR observations with a median seeing $\sim$ $1{^\prime}{^\prime}$ in the visual band. 
PRL's 1.2 m f/13 telescope is equipped with a Near-IR Camera \& Spectrograph (NICS) serving as one of the general purpose back-end instruments.  The wavelength bands covered are J, H, and Ks. The imaging detector is a Teledyne H1RG array with 1K x 1K pixels of 18 $\mu$m size each. In imaging mode, the instrument has a square FOV of $8^\prime$ x $8^\prime$ with a spatial scale of $0.5{^\prime}{^\prime}$ per pixel. The capabilities of the instrument were enhanced with the addition \citep{aarthy} of a polarimetric module (NICSPol), between the telescope optics \& NICS.  The technique of polarimetry is implemented using a 25.0 mm x 25.0 mm wire-grid polarizer (WGP) module at room temperature fixed in a motorised rotator originally used as a field rotator for a small telescope (Pyxis LE field rotator model from Pyxis Instruments). With this arrangement, the wire-grid serves as both a modulator \& an analyzer. The clear circular field of view at the image plane is $\sim$ $3.9^\prime$ in diameter due to vignetting by WGP. Images are taken at 4 angular positions ($0^{\circ}$, $45^{\circ}$, $90^{\circ}$ \& $135^{\circ}$) of the wire-grid polarizer.

\subsection{Observational procedure}
The observations of the L1340 with particular emphasis on the RNO 8 and surrounding area were fulfilled using NICSPol.  All observations were completed during dark nights, and at multiple epochs in November 2017 with appropriate polarisation standard stars done each night to facilitate the standardization procedure. 

The relatively small FOV of NICSPol (3.9$^{\prime}$ diameter) meant that multiple pointings were required to be observed for a complete coverage of the field.  We covered the region of L1340 from $\ell=$ 130.08$^{\circ}$ - 130.24$^{\circ}$,  $b=$ 11.44$^{\circ}$ - 11.60$^{\circ}$, an effective FOV of 9.6$^\prime$ x 9.6$^\prime$ with 6 positions (spread over different central coordinates) taken in succession. The images at 4 wire-grid position angles were acquired. With the individual exposure times of 50 sec, and three exposures per pointing per position angle, the effective exposure time was 150 sec per position angle per pointing.  Standard polarized and unpolarized stars were observed multiple times, in all filter bands, during the night, to get coverage of redundant standards over the night.

\section{Data reduction and analysis}
\label{3}

The observed data were reduced and analyzed using standard IRAF routines.
This involved a series of steps in the sequence of basic image reduction, astrometry, photometry \& polarimetry.
We created sky frames for each pointing by median combining all the rest of the observed pointings for a given WGP position. 
The minor shifts in the 4 object frames for different WGP position angles were corrected using the \textsc{imshift} task.\\
Astrometric solutions were fitted on the combined fits images using the \textsc{Astrometry.net} software\footnote{\url{http://nova.astrometry.net}}.\\ 
After fitting astrometric solutions, we used the 2MASS All-Sky Point Source Catalog for this region as input to the `sky2xy' command\footnote{part of the wcstools package} and obtained the image coordinates for further photometry. Photometry utilised \textsc{iraf} routines making use of aperture photometry \textsc{phot} \citep{stetson} followed by psf photometry with \textsc{pstselect, psf, \& allstar}.

\begin{table*}
\centering
\caption{Observational results of polarized standard stars. The standard value for \textit{p} $\&$ $\theta$ is taken from NOT $\&$ UKIRT links. The observed values are listed as \textit{p}$_{obs}$ $\&$ $\theta_{obs}$. The offset for position angles is given by $\theta_{off}$ }
\begin{tabular}{|c | c |c | c  | c | c | c | c |}
\hline 
\Tstrut\Bstrut  Star & Date & Filter & \textit{p}  &  $\theta$ & \textit{p}$_{obs}$  & $\theta_{obs}$ & $\theta_{off}$=$\theta_{obs}-\theta$ \\
\Tstrut\Bstrut           & & & (percent) & ($\circ$)     & (percent)   & ($\circ$)    & ($\circ$) \\ \hline
\Tstrut\Bstrut  HD204827 & 16/11/2017 & J &2.83 $\pm$ 0.07 & 61.1 & 2.83 $\pm$ 0.33 & -3.19 & -64.29 $\pm$ 6.55  \\
\Tstrut\Bstrut   & 16/11/2017 & H &  &  & 1.11 $\pm$ 0.49 & -1.19 & -62.29 $\pm$ 5.47 \\
\Tstrut\Bstrut             & 17/11/2017 & J & 2.83 $\pm$ 0.07  & 61.1 & 3.39 $\pm$ 0.49 & -16.72 & -77.82 $\pm$ 16.1 \\
\Tstrut\Bstrut             & 17/11/2017 & H & &  & 1.84 $\pm$ 0.33 & -4.94 & -66.04 $\pm$ 5.5\\ \hline
\Tstrut\Bstrut  HD283809       &  16/11/2017 & J & 3.81 $\pm$ 0.07 & 57 $\pm$ 1 & 3.26 $\pm$ 0.69  & -13.09  & -70.09 $\pm$ 14.5   \\ 
\Tstrut\Bstrut  &     16/11/2017 & H & 2.59 $\pm$ 0.07 & 58 $\pm$ 1 & 1.32 $\pm$ 0.69  & -6.27 & -64.27 $\pm$ 7.2 \\ \hline

\end{tabular}
\label{observations_1}
\end{table*}

\subsection{Polarisation calculation}
The measured magnitudes were converted to flux units for polarisation calculations. 
After obtaining the intensity measurements of the stars at each position angle of the WGP, the polarimetric analysis was carried out using the Stokes method by applying the formulae below:

\begin{equation}
    I = \frac{F_{0}+F_{45}+F_{90}+F_{135}}{4}
    \label{I}
\end{equation}
\begin{equation}
       Q = \frac{F_{0}-F_{90}}{2}
       \label{Q}
\end{equation}
\begin{equation}
      U = \frac{F_{45}-F_{135}}{2}
      \label{U}
\end{equation}

\noindent where $F_0$, $F_{45}$, $F_{90}$, and $F_{135}$ are the fluxes at 0$^\circ$, 45$^\circ$, 90$^\circ$, and 135$^\circ$ position angles of the WGP.\\
Using the Stokes vectors, I, Q \& U, we derived the degree of polarisation  (P) \& position angle of polarisation ($\theta$), as formulated below:
\begin{equation}
      P = \frac{\sqrt{Q^{2}+U^{2}}}{I}
      \label{PF}
\end{equation}
\begin{equation}
      \theta = \frac{1}{2}tan^{-1}\frac{U}{Q}
      \label{theta}
\end{equation}
The error estimates for the P \& $\theta$ were derived using the fundamental error propagation algorithms:

\begin{equation}
      \sigma_{P} = \frac{1}{I}\sqrt{\frac{Q^{2}{\sigma_{Q}}^{2} + U^{2}{\sigma_{U}}^{2}}{Q^{2}+U^{2}} + \frac{Q^{2}+U^{2}}{I^{2}}{\sigma_{I}}^{2}}
      \label{sigmaPF}
\end{equation}
\begin{equation}
      \sigma_{\theta} = \frac{1}{2}\sqrt{\frac{Q^{2}{\sigma_{U}}^{2} + U^{2}{\sigma_{Q}}^{2}}{(Q^{2}+U^{2})^2}} rad
      \label{sigmatheta}
\end{equation}

\subsection{Polarisation calibration}

The data are calibrated with observed polarisation standards i.e. unpolarised \& polarised standard stars. The unpolarised stars are used to account for any instrumental polarisation if present in the system.  Since the NICSPol module is the first element after the telescope and before any asymmetric reflection in the instrument, we do not expect any instrumental polarisation. The results of the standard star observations are discussed in the instrument paper of NICSPol \citep{aarthy}. The instrumental polarisation is close to 1\% uncertainty in the polarisation fraction. The observation of polarised standard stars allows for the conversion of our polarisation position angles to celestial coordinates (table \ref{observations_1}). Several standards were observed in filters (J, H) on the multiple nights of the observing runs.  The standards were taken from the lists maintained by UKIRT \footnote{\url{https://about.ifa.hawaii.edu/ukirt/calibration-and-standards/
unpolarized-standard-stars/.}} \&  Nordic Optical Telescope \footnote{\url{http://www.not.iac.es/instruments/turpol/std/hpstd.html}}. 

\subsection{IR polarisation values and large errors}
\label{largeIR}
All the stars from our observation and analysis satisfy the criteria of $P$/$\sigma_{P}$ $>$ 3, and the degree of polarisation $P$ $<$ 30$\%$.  We have dropped sources with $P$ $>$ 30$\%$ since they were all faint or were at the edge of the circular field of NICSPol and the photometry would not be reliable. 

We present a data set with a larger degree of polarisation  than theoretically expected to arise from purely interstellar extinction at the IR wavelengths \citep{Jones1989}, for some of the stars.  
However, we note that at the individual source level, similar numbers are seen in the case of MIMIR \citep{mimir} and SIRPOL \citep{Hatano2013} data sets in other directions. In the case of stars sitting in or towards reflection nebulae, one may see a much higher degree of polarisation \citep{Jones1989}. 
The large values may be attributed to a few observational limitations in our case: non-simultaneity of images at 4 angles, lower SNR, and significant  background from the warm WGP.  Of all these points, it appears that the varying infrared sky background (over the period of the observations of the 4 WGP angles) may be the most significant contributor to the uncertainty in the photometry in the individual frames.  
It has been mentioned in \cite{ss} that with low S/N, the derived polarisation values tend to be overestimated. This would be especially applicable for the fainter stars. A correction has been suggested by them, for reducing the overestimated polarisation values, expressed as
\begin{equation}
    P_{corr} = \sqrt{(P_{o}^{2}) - \delta P^{2}}
\end{equation}
where, $P_{o}$ is the observed degree of polarisation in our results.
We considered the above correction and found that it did not make a significant change in our results and hence do not include this correction in the final tables and figures presented in this work.

\section{Supplementary Data}
\label{4}

\subsection{\label{sec:2massgaia}2MASS and Gaia} 
\begin{figure}
    \centering
    \includegraphics[width=\columnwidth]{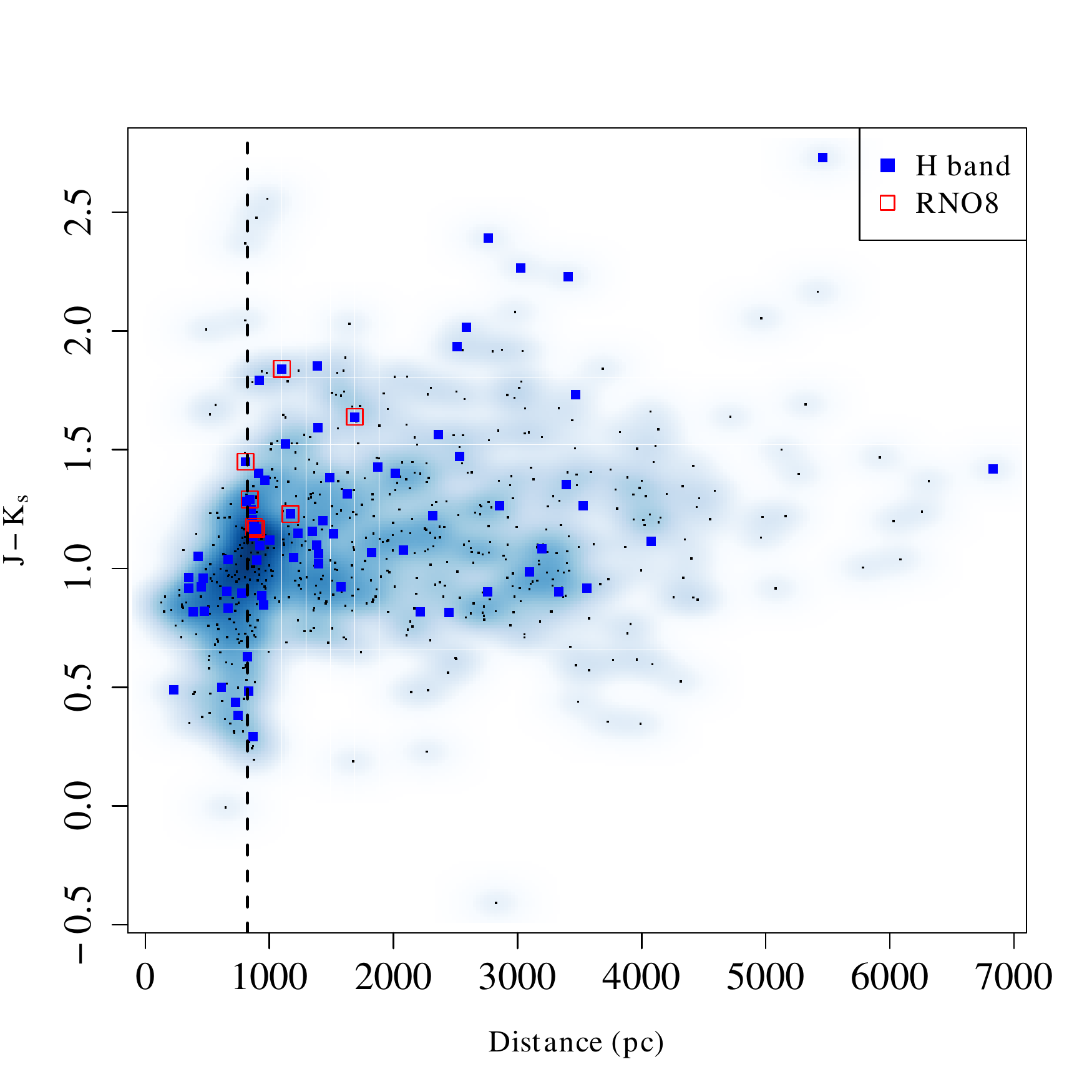}
    \caption{2MASS $J-K_s$ color vs Gaia distance.  Blue-filled squares show the stars for which polarisation is measured in the H band. Stars of the RNO 8 area are shown as red open squares. A vertical dashed line is marked at 825pc - denoting the distance to the cloud as per \citet{Kun}. }
    \label{distJK}
\end{figure}

\begin{figure}
    \includegraphics[width=\columnwidth]{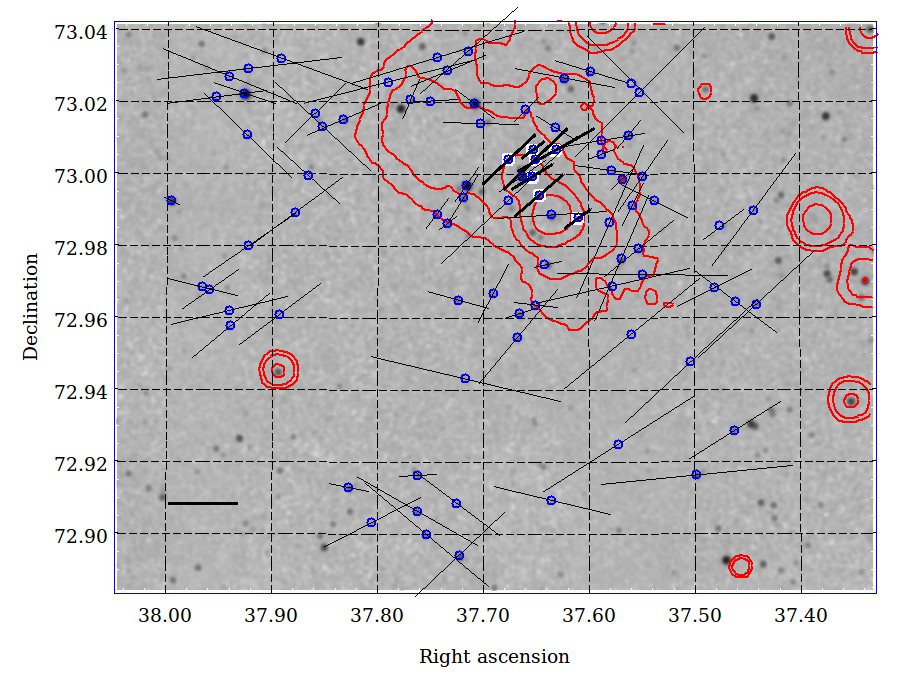}
    \caption{H band polarisation vectors (thin solid lines) plotted on the H filter 2MASS image of L1340. The thick black vectors are for stars corresponding to the RNO 8 clump, a small nebulous cluster within the L1340 cloud. We emphasize the highly ordered pattern of polarisation vectors for the stars within this clump. The solid line at the bottom left corner indicates a 10\% degree of polarisation value.  The contours in red are from the WISE (band 4) image of the same area.}
    \label{allpol}
\end{figure}

The 2MASS and Gaia data have been used in this study to understand the variation of polarisation with NIR color and with distances along the line of sight. The 2MASS All-Sky Catalog for Point Sources \citet{2MASS}, has a total of 765 sources detected in the 10$^{\prime}$ search radius for the L1340 cloud. \\\\
The Gaia EDR3 data
\citep{baileredr3,gaia2016b, gaia2020a} have been downloaded from the Vizier\footnote{http://vizier.u-strasbg.fr/} database. The distance estimates from inverting the parallax have been dealt with in a self-consistent manner in this data release, using a Bayesian Inference approach. 
A total of 1003 Gaia sources were detected in the 10$^{\prime}$ search radius centered on LDN 1340. Of these, 686 sources are seen matched in both Gaia and 2MASS.\\
The recently released Gaia DR3 data \citep{gaia2022, 2022j} was also utilised for further analysis which included the proper motion data and extinction values A$_G$ for the stars within the L1340 cloud.\\
Figure \ref{distJK} uses this combined catalog to show the colour vs distance diagram for all the 2MASS and Gaia sources. Figure \ref{allpol} shows the H band image (2MASS) of the field covered by our NICSPol observations in the equatorial coordinate system.  The 2MASS image of the L1340 region was used after mosaicing individual 2MASS tiles using \textsc{montage} tool. Our coverage of the field with NICSPol is incomplete due to the circular fields of the individual NICSPol tiles as discussed in the previous section. In both Figures (\ref{distJK}, \ref{allpol}), the individual stars measured for polarisation are shown by separate symbols and will be discussed in section \ref{sec:results}.

\subsection{WISE data}
\begin{figure}
    \includegraphics[width=\columnwidth]{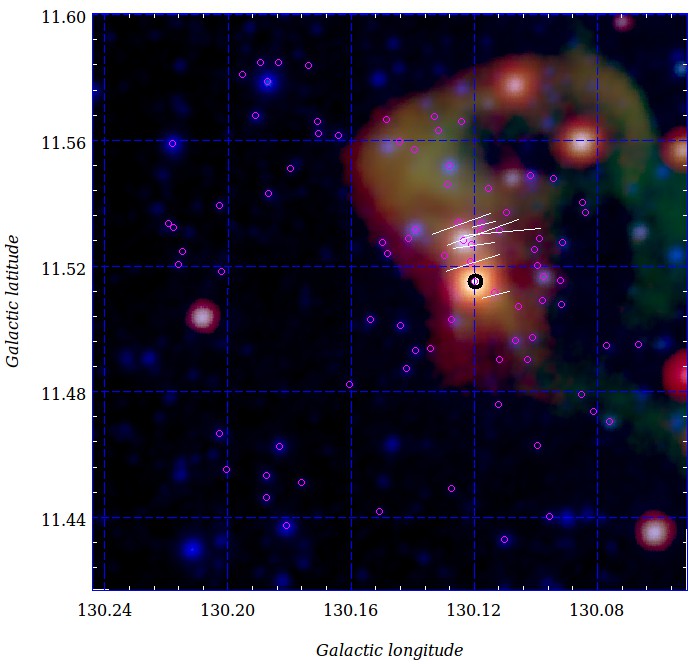}
    \caption{H band polarisation vectors overlaid for the RNO 8 stars on the RGB colour composite of the WISE 4,3,2 bands.  The black circle marks the core of RNO 8.  Further discussion is in the text.
    }
    \label{allpolwise}
\end{figure}
The WISE mission \citep{2010AJ_Wright_WISE} maps the interstellar dust over the whole Galaxy.  This is from the presence of PAH emission features in the W1 and W3 filters, 3.4 $\mu$m and 12 $\mu$m respectively.  The other two filters, W2 and W4 (4.6 $\mu$m and 22 $\mu$m respectively) measure the continuum emission from the grains (small and large).  We note that the 22$\mu$m filter is expected to see both `stochastic emission from small grains and the Wien tail of thermal emission from large grains' \citep{2010AJ_Wright_WISE}.\\
The WISE band images for the L1340 cloud at 3.4 $\mu$m, 4.6 $\mu$m, 12 $\mu$m, and 22 $\mu$m were imported into {\sc{ds9}} 
\footnote{Data downloaded from \url{https://irsa.ipac.caltech.edu/Missions/wise.html}} in RGB mode.  The colour scaling and levels were chosen to bring out the dust features. This nebulosity is very nicely evident around the RNO 8 area (marked by a black circle in Fig. \ref{allpolwise}) in the RGB colour-composite (with W2, W3 and W4 represented as blue, green, and red colours). The 22 $\mu$m band was also used to mark contours over the cloud (Fig \ref{allpol}), revealing the clumpy structure of the dust towards and in the L1340 region. 

\section{Results \& Discussion}
\label{sec:results}

In the region covered by our NICSPol observations, we have 84 stars in the  H band. We cross-identified the polarisation measurements from these stars with the 2MASS,  Gaia EDR3 and DR3 surveys to get proper distance information and thence characterize them.  In total, we have 76 stars in the H band with 2MASS and Gaia counterparts. The cross-match of the different surveys was done in \textsc{TOPCAT} with the Sky algorithm, using a search radius of 2${^\prime}{^\prime}$. 

\begin{table*}
\setlength\tabcolsep{4pt}
\caption{
NICSPol linear polarization measurements (p$_{H}$) \& polarization position angle ($\theta$) of the RNO 8 stars. The H magnitudes and distances to individual stars from 2MASS \& Gaia survey data are also included.  The central star of RNO 8 is shown in bold in this table and discussed further in the text. 
}
\centering
\begin{tabular}{| c | c | r | r | r | r | c | c |}
\hline
  \multicolumn{1}{|c|}{RA} &
  \multicolumn{1}{c|}{DEC} &
  \multicolumn{1}{|c|}{p$_{H}$} &
  \multicolumn{1}{c|}{$\theta_{H}$} &
  \multicolumn{1}{c|}{Hmag} &
  \multicolumn{1}{c|}{rgeo} &
  \multicolumn{1}{c|}{Proper} &
    \multicolumn{1}{|c|}{SIMBAD}\\
  \multicolumn{1}{|c|}{(deg)} &
  \multicolumn{1}{c|}{(deg)} &
  \multicolumn{1}{c|}{(percent)} &
  \multicolumn{1}{c|}{($\circ$)} &
  \multicolumn{1}{c|}{(mag)} &
  \multicolumn{1}{c|}{(pc)}  &
  \multicolumn{1}{c|}{motion} &
  \multicolumn{1}{|c|}{identification}\\
\hline
%
37.67549 & 73.00384   &  10 $\pm$   1 & 133$\pm$5   & 12.94 $\pm$ 0.03 & 893$^{935}_{848}$ & 2.4 & YSO \\[0.2cm]
37.65248 & 73.00647 &	4 $\pm$   1 & 127$\pm$7  & 13.14 $\pm$ 0.03  & 808$^{848}_{765}$ & 2.3 & YSO\\[0.2cm]
37.65338 & 72.99905 &	7 $\pm$   1 &122$\pm$3 & 12.51 $\pm$ 0.04 &  876$^{906}_{850}$ & 2.2 &  T Tauri\\[0.2cm]
37.60990 & 72.98741 &	5 $\pm$   1 &127$\pm$7 & 13.71 $\pm$ 0.05 & 1101$^{1450}_{884}$ & 2.1 & -\\[0.2cm]
37.65020 & 73.00396  &  13 $\pm$   3 &134$\pm$5 &  14.76 $\pm$ 0.07 & 1170$^{1598}_{948}$ & 2.3 & -\\[0.2cm]
37.63104 & 73.00655  &  13 $\pm$   2 &119$\pm$5 & 14.90 $\pm$ 0.06 &  1689$^{2246}_{1134}$ & 3.0 & YSO\\[0.2cm] 
37.64659 & 72.99383 &   9 $\pm$   1 &131$\pm$4 & 13.50 $\pm$ 0.03 & 842$^{918}_{769}$ & 2.3 & -\\[0.2cm] 
\textbf{37.63532} & \textbf{72.98826} & \textbf{16 $\pm$   1} & \textbf{94$\pm$1} & \textbf{12.23 $\pm$ 0.03} & \textbf{5458$^{8269}_{3812}$} & 2.3 & \textbf{T Tauri}\\[0.1cm]
\hline
 \end{tabular}
 \label{RNO8_table}
 \end{table*}

\subsection{Color \& distance information}
Figure \ref{distJK} shows the distribution of the 2MASS colours and the corresponding Gaia distances. The distance to the cloud is quoted to be 825$^{+110}_{-80}$ pc from the photometric and spectroscopic study of L1340 by \citep{Kun}. We show this by a dashed line in the figure.  
\begin{figure}[H]
    \centering
    \includegraphics[width=\columnwidth]{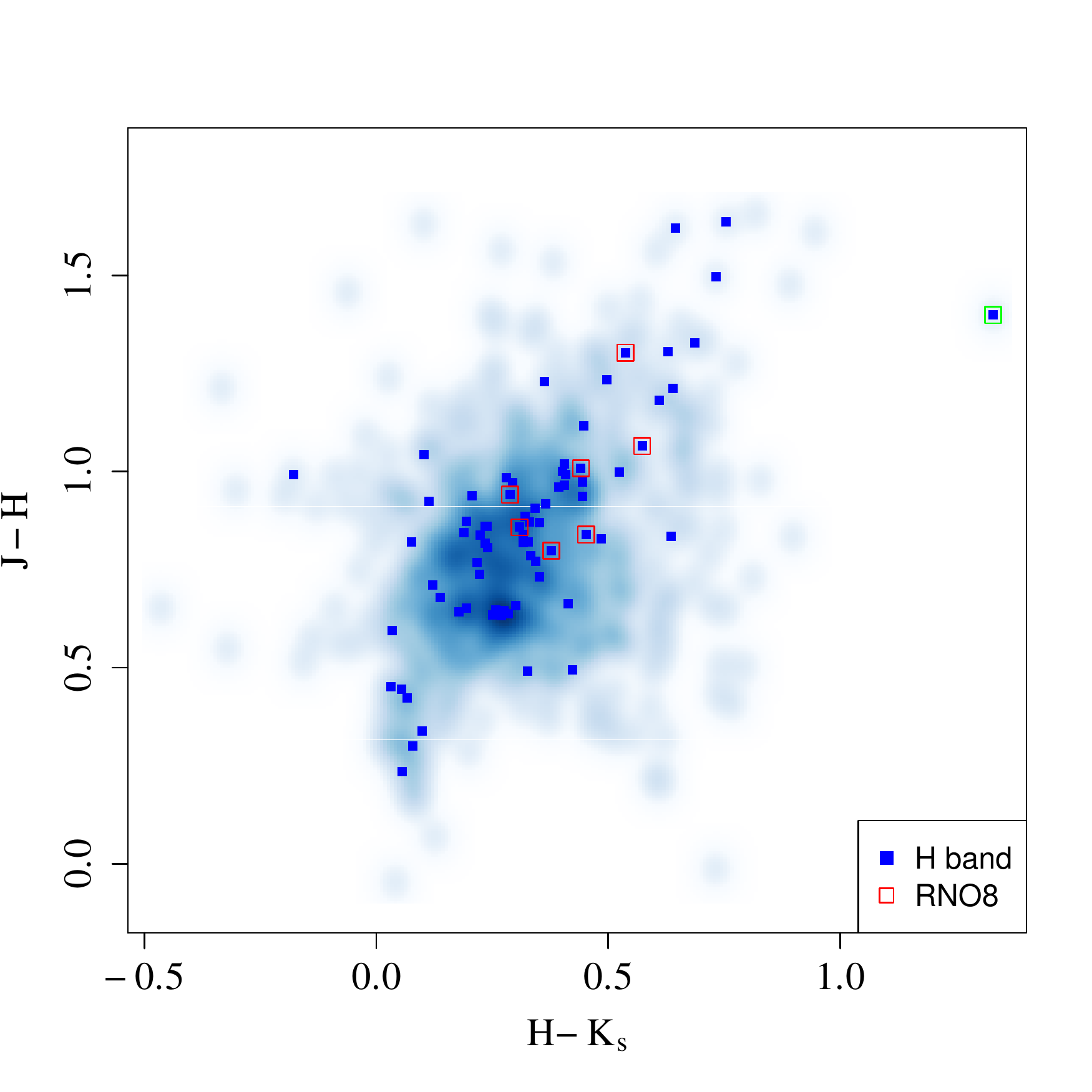}
    \caption{Color-color diagram using 2MASS colors for the stars in L1340. The stars detected in the H band with NICSPol are marked with blue squares over the smoothScatter density plot of the 2MASS stars. The red open squares mark the stars of RNO8 cluster, and the green open square is for the central star of RNO8.}
    \label{ccdia}
\end{figure}

We see that this estimate is quite robust, since there is a sharp increase in the number of sources, along with a reddening in colour, beyond this distance.
The 2MASS two-colour diagram (CCD) is shown in Figure \ref{ccdia}. The stars observed by NICSPol are shown as blue squares over a `smoothScatter' distribution of the 2MASS detections towards the L1340 cloud. The RNO8 stars are  overlaid as large red squares. From the color-color diagram it is evident that these stars are redder compared to the other stars in the CCD.

\subsection{NIR Polarimetry}
The position angles of the stars in the RNO 8 group show a preferential direction along the Galactic plane, visible from their alignment nearly parallel to the grid lines in Figure \ref{allpolwise}, Table \ref{RNO8_table}. 

\begin{figure}[H]
    \centering
    \includegraphics[width=\columnwidth]{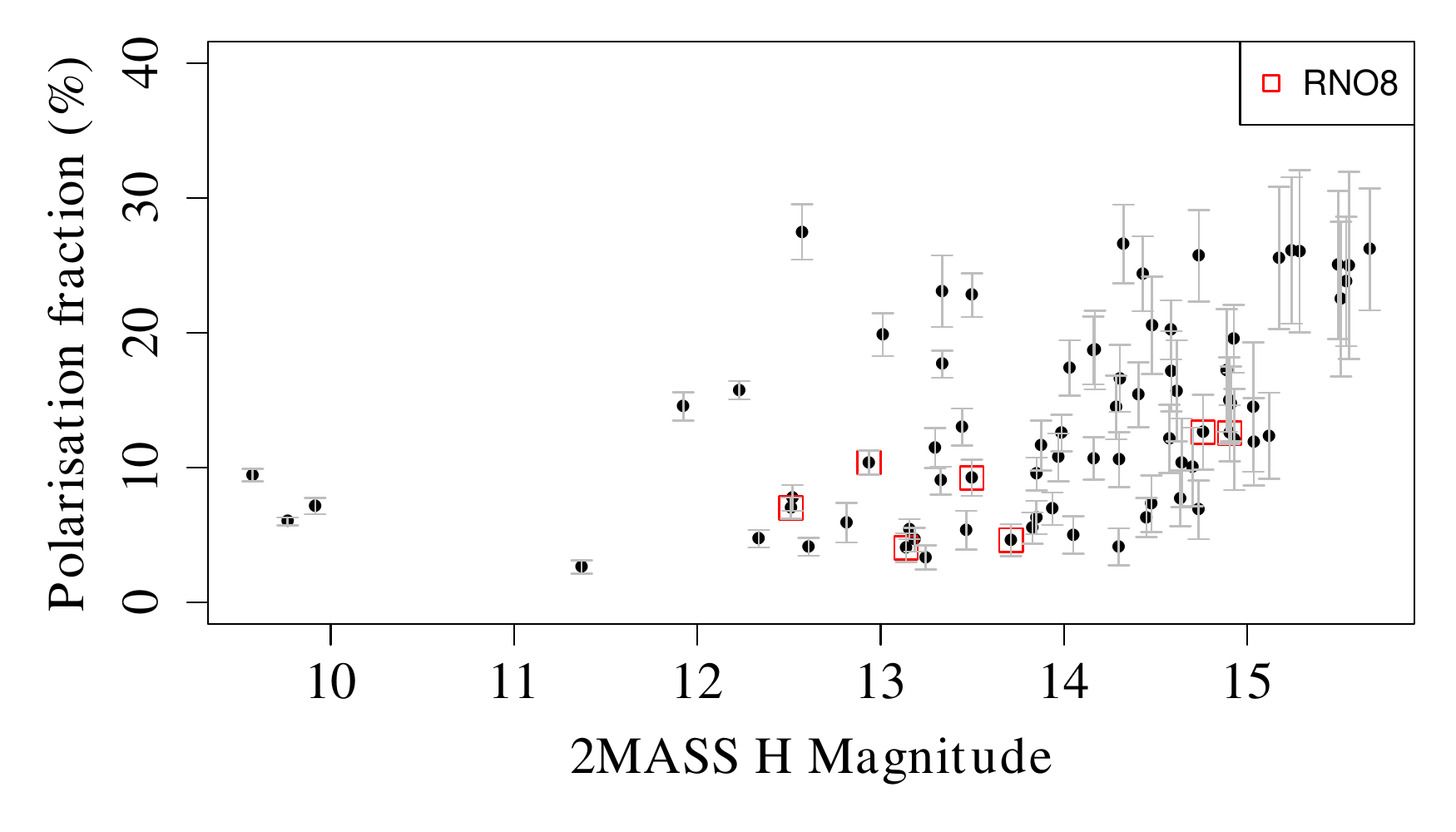}
    \caption {Polarisation in the H band as a function of the corresponding 2MASS H magnitudes.}
    \label{magPF}
\end{figure}
The polarisation values are plotted as a function of the 2MASS H band photometry in Figure \ref{magPF}.
The polarisation values and the corresponding uncertainties for the fainter stars have been discussed in section \ref{largeIR}.  
To investigate the orientation of the magnetic field in the direction of the L1340 molecular cloud, we created linear polarization maps for the H wavelength band. The map in Figure \ref{allpol} shows a vector plot overlaying the polarisation vectors (degree of polarisation \& position angle) for the H band on the 2MASS H filter image. The overall orientation of most of the polarisation vectors is nearly aligned with the Galactic plane. There are large departures in some of the stars away from the nebulosity marked by the 22 $\mu$m WISE band 4 contours.\\ 

\begin{figure}[H]
    \centering
    \includegraphics[width=0.9\columnwidth]{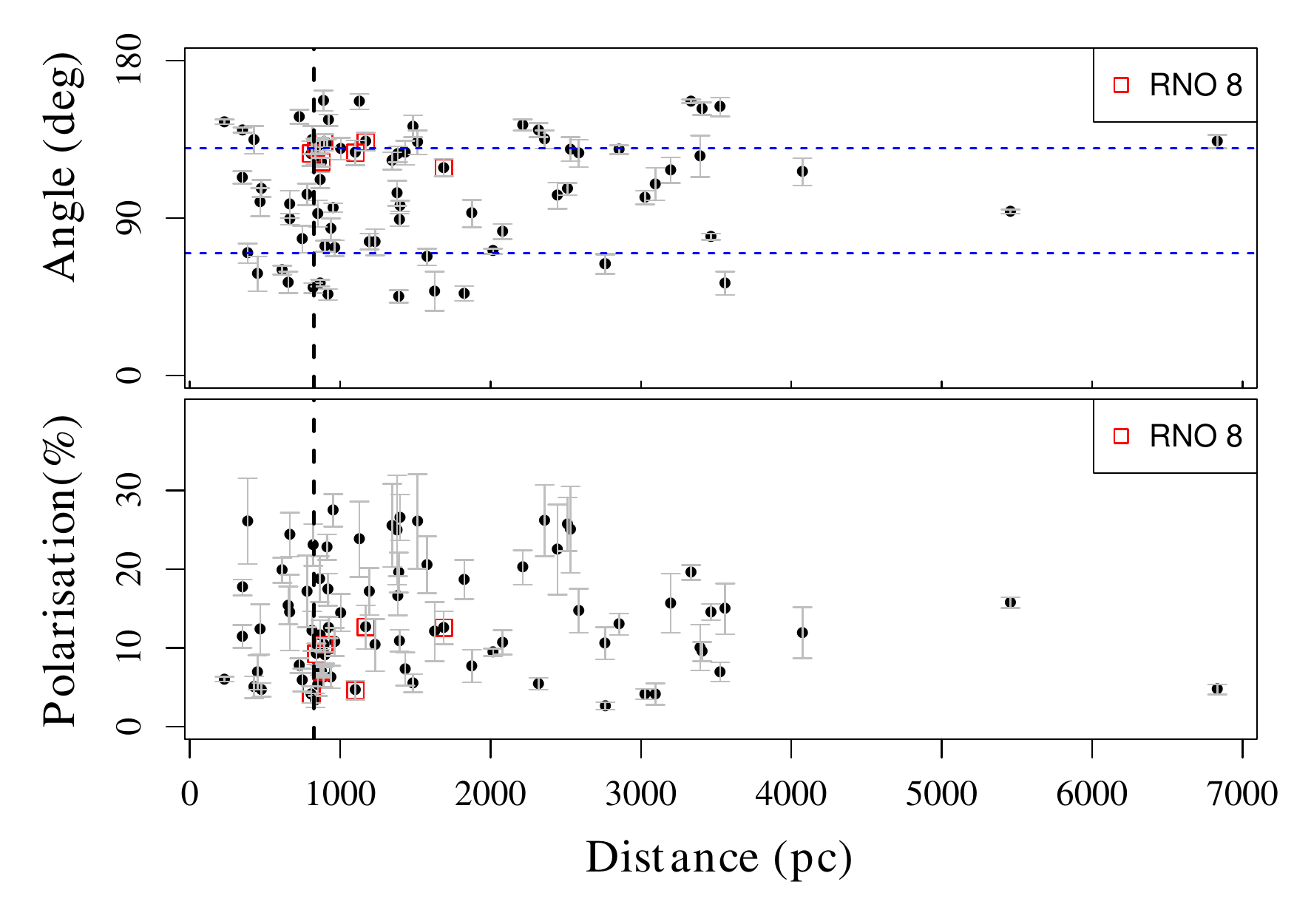}
    \caption{Distance (in pc from Gaia) vs polarisation percentage (lower panel) and polarisation position angle (upper panel) for H band measurements.  Stars in the vicinity of RNO 8 are shown with larger symbols.  }
    \label{distPF}
\end{figure}

\begin{figure}[H]
    \centering
    \includegraphics[width=0.9\columnwidth]{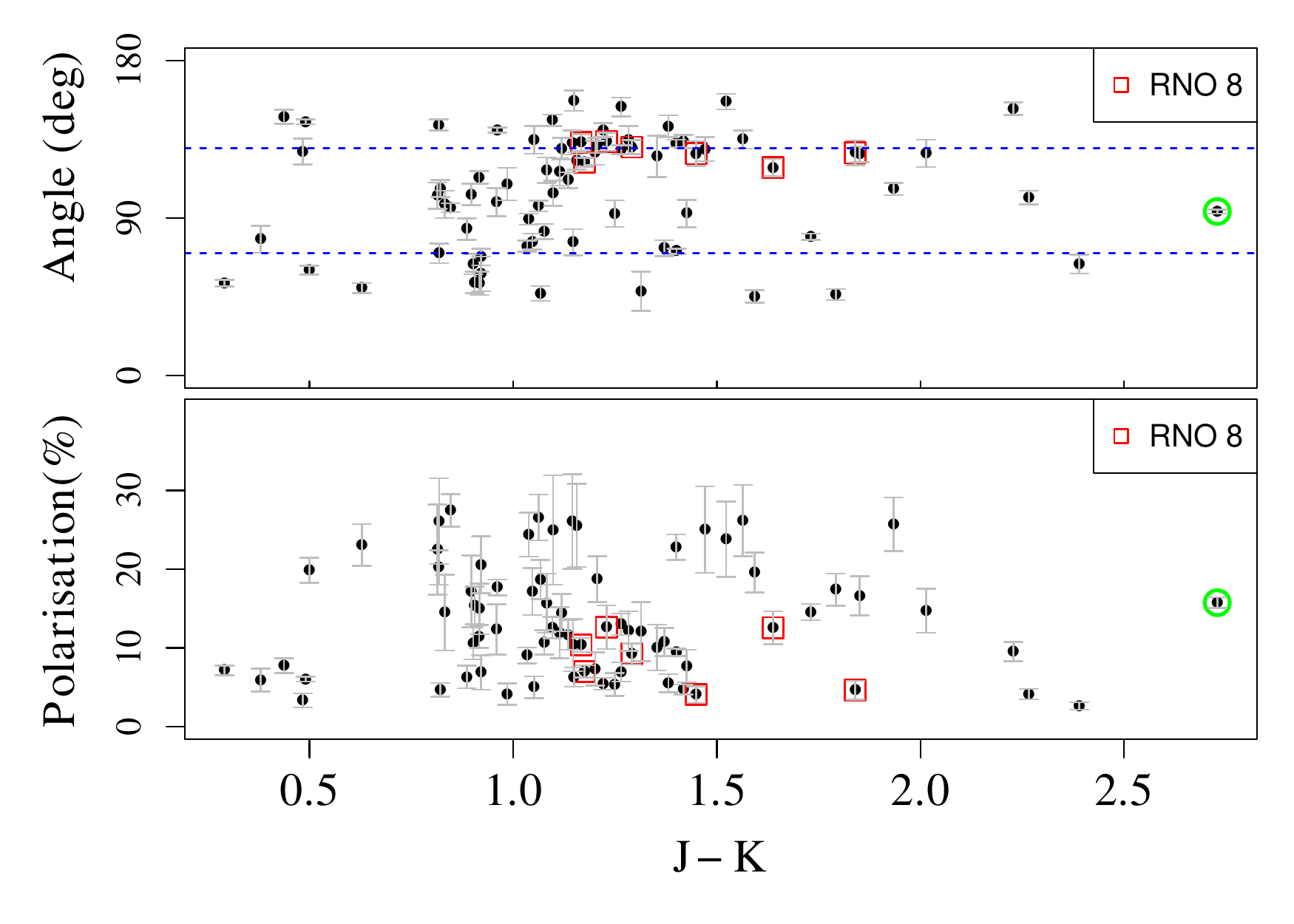}
    \caption{Polarisation in the $H$ band (percentage in lower panel and angle in upper panel) vs $J-K_s$ colour (from 2MASS). The central star of RNO 8 is shown with a green open circle.}
    \label{JKpfpa}
\end{figure}

\begin{figure}[H]
    \centering
    \includegraphics[width=0.4\textwidth]{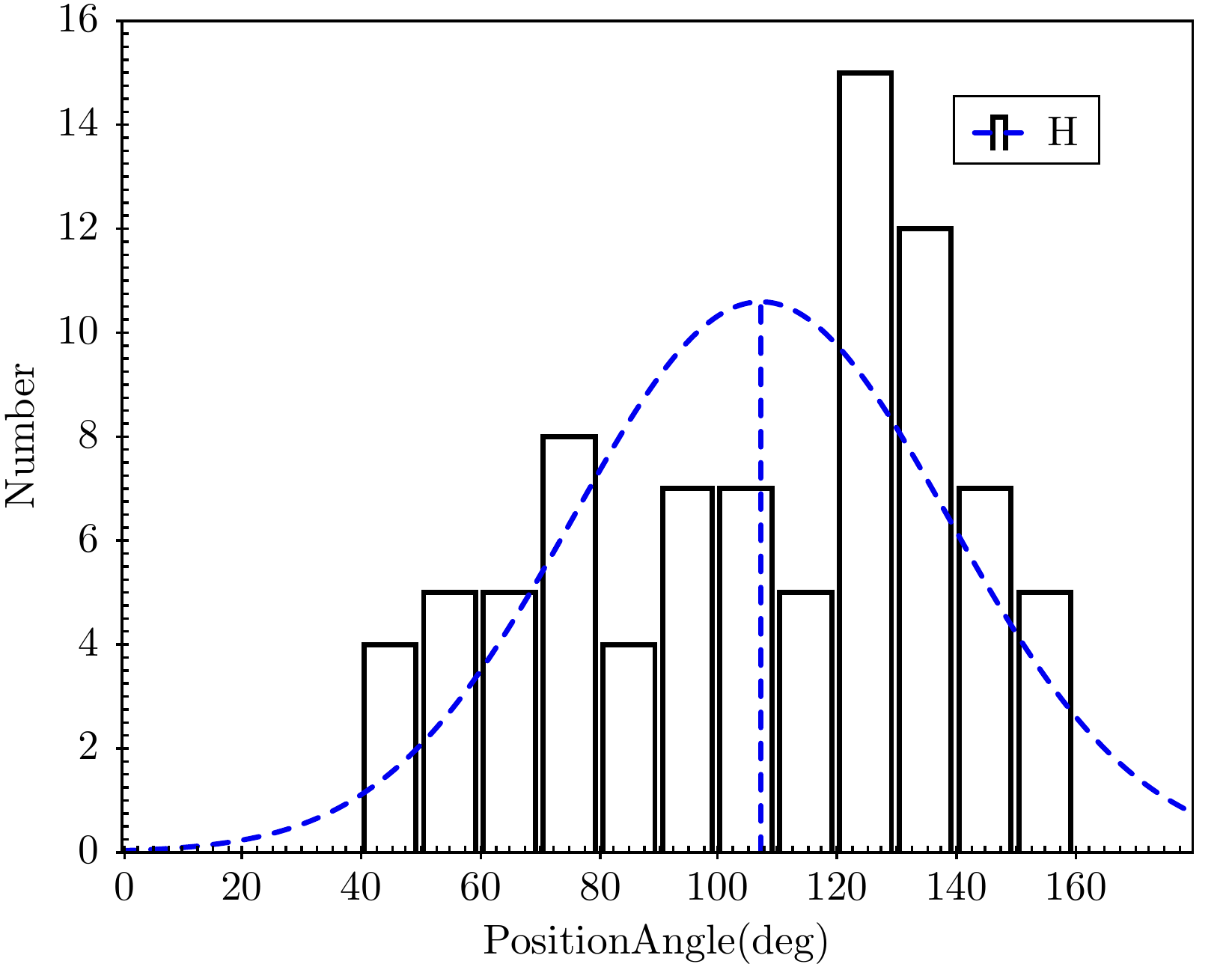}
    \caption{Position angle distribution for the stars was measured in the H band. A gaussian fit is overlaid on the panel. Two dominant angles could be seen from the plot.}
    \label{angleHist}
\end{figure}

The distribution of the position angle with distance for the H band is plotted in Figure \ref{distPF}. Two dominant angles (PA $=$ 130$^\circ$ and 70$^\circ$) are seen in the H band histogram (Fig. \ref{angleHist}). These angles are marked in the position angle panels of Figures \ref{distPF} and \ref{JKpfpa} as horizontal lines. For the stars located at the distance of the L1340 cloud, there is a large scatter in the polarization position angle covering the extent of the two dominant angles.\\

\begin{figure}[H]
    \centering
    
    \includegraphics[width = 0.45\textwidth]{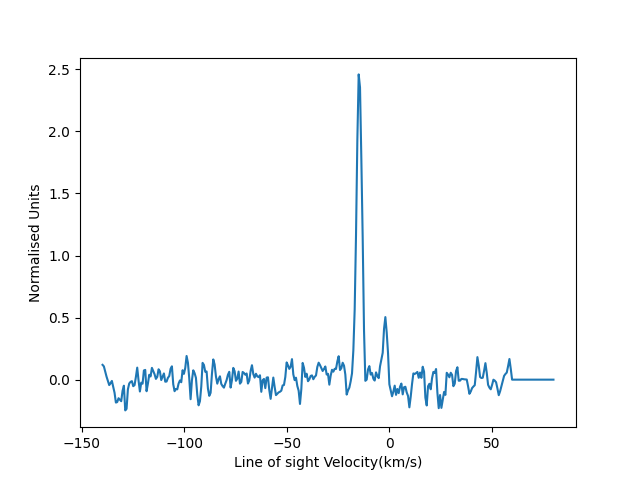}
    \caption{Line of sight velocity plot from CO data are taken from \citet{dame} composite survey.}
    \label{COspec}
\end{figure}
The plot in Figure \ref{COspec}, shows the line of sight velocity (V$_{lsr}$) for the Lynds cloud. The distinctive peak seen is at a velocity of -14.92 km/s, and the comparatively smaller peak is at a velocity of -1.944 km/s. Similar velocity has been quoted for the L1340 cloud in \cite{Kun1994}, using CO spectra data taken in the field containing LDN 1340 from the Nagoya university 4 m telescope. The characteristic peak tending to a radial velocity of -14.5 km/s was estimated. Using \cite{reid}\footnote{\url{http://bessel.vlbi-astrometry.org/revised_kd_2014}}, we could estimate the distances to these V$_{lsr}$. The distance corresponding to the taller peak comes out to be 0.82$^{0.51}_{0.49}$ kpc. Unfortunately, the distance to the smaller peak couldn't be estimated from the V$_{lsr}$ using the same code.

\subsection{The RNO 8 cloud core}
\begin{figure}[H]
    \centering
    \includegraphics[width=0.9\columnwidth]{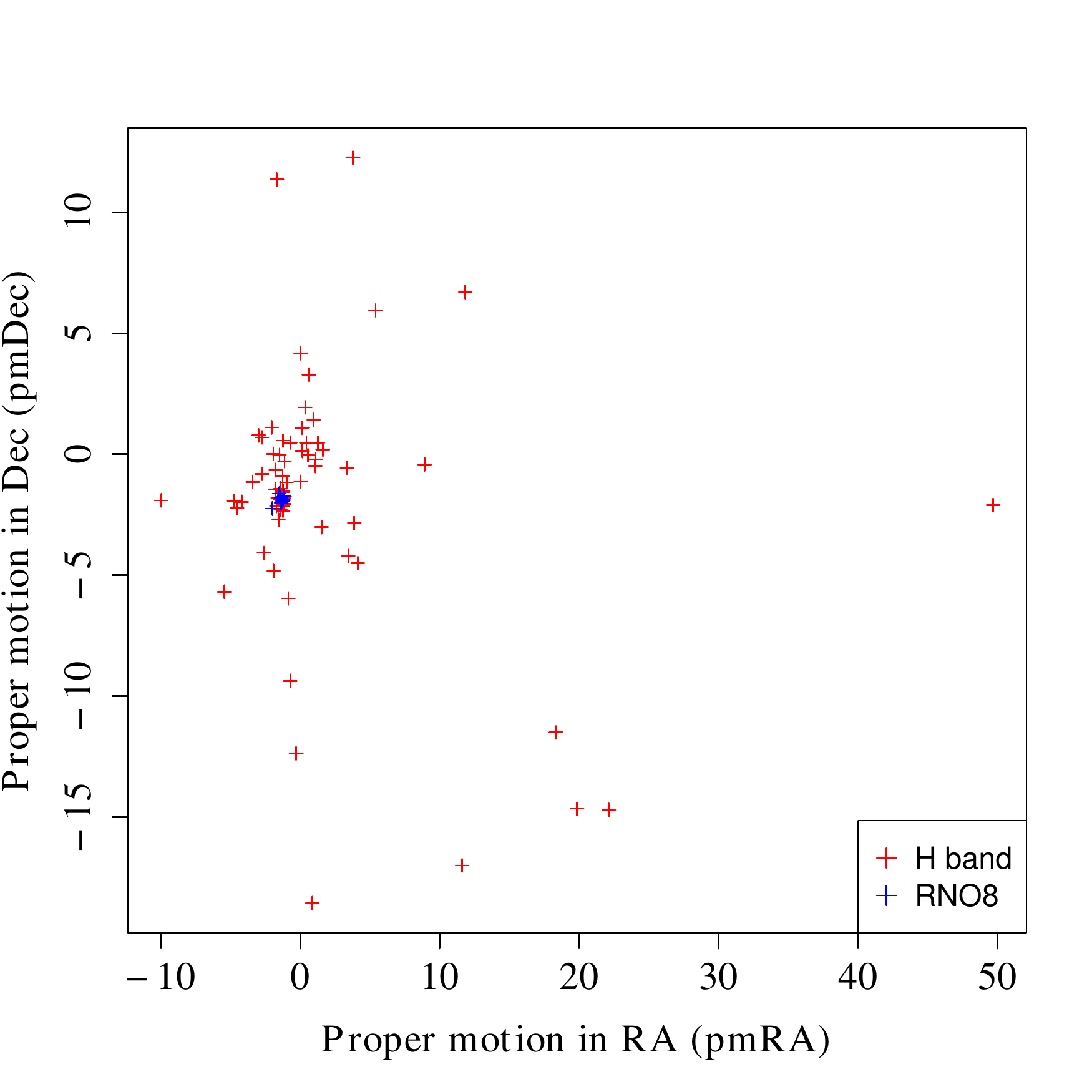}
    \caption{Proper motion in the stars belonging to the L1340 cloud. The red cross is for all the stars whose polarisation is quoted in the H band. The RNO 8 clump stars are shown in the blue cross. }
    \label{pmplot}
\end{figure}

The Figures \ref{allpol} \& \ref{allpolwise} show the RNO 8 stellar group between $\alpha$ = 37.60$^{\circ}$ - 37.70$^{\circ}$, $\delta$ = 72.98$^{\circ}$ - 73.00$^{\circ}$; $l$ = 130.12$^{\circ}$ $\&$ $b$ = 11.51$^{\circ}$. These are red and nebulous objects with associated groups of fainter stars \citep{cohen,RNO1}. From the proper motion plot in Fig.\ref{pmplot}, it is clear that the stars in this clump share a similar proper motion and hence are considered member stars for the RNO8 clump for the analysis in this paper.
The star identified by \citet{Kun} at the core of the RNO 8 nebulosity ($\alpha$ = 37.635 $\&$ $\delta$ = 72.988) is at a distance of 5~kpc as per the corresponding value from the Gaia data.  The RNO 8 region is well studied photometrically \& spectroscopically by Kun et al., and one major result is that the central star is a candidate embedded intermediate-mass young star of the optical nebulosity RNO 8. They suggest it to be a young star of 2 - 2.5 M$_{\odot}$ \citep{Kun}. Based upon the `renormalised unit weight error' $ruwe$ parameter, whose threshold value for reliable astrometry is 1.4, the central star has $ruwe$ $>$ 1.4, which makes the Gaia distance data unreliable for this star.\\
Table \ref{RNO8_table}, lists the polarisation values for the stars of the RNO 8 group along with the corresponding 2MASS H magnitudes and the distance and proper motion information from Gaia.

\subsubsection{Polarisation distribution with distance \& colour:}
\label{sec:pf}

An indirect estimation for distances to the cloud can be devised from the polarimetric data. A uniform distribution of material along the line-of-sight will exhibit an increasing trend in extinction, from the relation   $<A_{v}/{L}>$ $\approx$ 1.8 mag kpc$^{-1}$ \citep{whittet}. The presence of a cloud will enhance the number density of the material in the line-of-sight from which the background stellar light is passing. The starlight reaching us, travelling through a medium will see an increment in the degree of polarisation at the distance to the cloud because of uniform dust presence.  \citep{eswaraiah}.
 
Looking at the lower panel in Figure \ref{distPF}, it seems that the stars part of RNO 8 (marked with red open squares) for which polarization is measured show an increase in the degree of polarisation with the distance in the cloud, assuming that the cloud starts just before the dotted line (at 825~pc) in the figure. But because of the statistics, we cannot visibly conclude the increment in polarisation degree to be significant in our case. These stars also have consistent angles, with very small dispersion, as seen in the upper panel of Figure \ref{distPF}.  We interpret this to mean that within the RNO 8 cloud core the magnetic field orientation remains uniform. 
In Fig \ref{JKpfpa}, we have plotted the 2MASS $J-K_s$ colour vs degree of polarisation in the lower panel.  The upper panel shows the variation of position angle.  We see that the stars in RNO 8 (marked by red open squares) do not show significant variation in the angle of polarisation over a large range in colour.  These  stars are identified as YSO and T Tauri (SIMBAD identification - see Table \ref{RNO8_table}) in the work by \citet{Kun}.  
Since we do not see any variation in position angle for these stars (apart from the star in the direction of the center of the core) it appears that they do not affect the overall magnetic field orientation despite being embedded in the dust.\\

\subsubsection{Estimation of magnetic field strength:}
The plane of the sky magnetic field can be calculated from the inclusion of polarimetric observations in addition to other parameters known from photometric data. The magnetic field strength as formulated by Chandrashekhar and Fermi (CF model) \citep{CF}, depends on the mean local density of the cloud $\rho$, the line-of-sight velocity dispersion $\delta v$, and the dispersion in the position angle of polarisation $\delta\theta$. The equation can be expressed as,
\begin{equation}
    B_{p}=\mathcal{Q} \sqrt{4 \pi \rho} \frac{\delta v_{\operatorname{los}}}{\delta \theta}
\label{mag}
\end{equation}
\citep{ostriker} has performed numerical simulations for the estimation of magnetic field strength and compared it with the CF model. Based on his results, he suggested a value $\mathcal{Q}$ $\sim$ 0.5, for angle dispersion $<$ 25$^{\circ}$. 
The values for $n(H_{2})$ and velocity dispersion based upon the observations of $C^{18}O$ from the earlier work of \citet{Kun1994} on this cloud will be used.  They have quoted velocity dispersion and mean cloud density for three cores A, B, and C within the L1340. Based on the RNO 8 coordinates, we use the numbers for core B of the cloud, which contains the RNO 8 cluster. The estimated value for $n(H_{2})$ = 830 $cm^{-3}$, and $\Delta v$ = 0.9 $kms^{-1}$. Using these values with our position angle dispersion of $5.17^{\circ}$, the $\Vec{B}$ strength in the plane of sky is calculated as $\approx$ 42 $\mu G$ for the RNO 8 core. This value is in accordance with the magnetic field strength $\sim$ 20 - 200 $\mu G$ in other molecular clouds \citep{Kwon_2016}.
 
The Gaia distance to the star in the direction of the center of the RNO 8 core is highly uncertain.  However since it is identified to be a T Tauri (see the star identified in bold font in Table \ref{RNO8_table}) star, we may consider it to be a part of the cloud complex.  Including the measurement of polarisation for this star increases the dispersion in the polarisation position angle to 12$^{\circ}$. From equation \ref{mag}, the calculated $\Vec{B}$ strength is found to be 18 $\mu G$. 
In such a case, the mean magnetic field strength is decreased with the increase in the dispersion of the magnetic field vector direction due to varying cloud structure (or multiple clouds) in the line of sight giving rise to different position angles in polarisation. The consequences of dispersion in the magnetic field direction have been discussed in \cite{ostriker}.

\section{Conclusions}
\label{6}
Near-Infrared polarisation in the H band over the L1340 molecular cloud has been presented for a FOV of 9.6$^{\prime}$ x 9.6$^{\prime}$ using PRL's 1.2 m telescope with the NICSPol instrument.  We conclude a few important results for the RNO 8 region in particular. The results are :
\begin{itemize}
    \item[1.] NIR polarimetry was performed on a region within L1340, and measurements for a total of 76 stars are reported in the H band along with their 2MASS photometry and Gaia distance and other astrophysical parameters where available.
    \item[2.] The distribution of the position angle of polarisation is very consistent and is almost aligned with the Galactic magnetic field, which implies that the dichroic extinction causing the polarisation due to the dust grains being aligned with the magnetic field permeating the galaxy.
    \item[3.] The magnetic field strength for the RNO 8 region within the L1340 cloud was estimated using the Chandrashekhar-Fermi method. A 42 $\mu$G magnetic field was estimated which is similar to the strength present in other such areas. 
    \item[4.] Considering the star in the direction of the center as part of the RNO 8 core, reduces the magnetic field strength to $\sim$ 18 $\mu$G, due to the increase in position angle dispersion. \\
\end{itemize} 
    
One important conclusion from the work presented here is that the use of single beam polarimetry is not very effective at measuring the relatively low interstellar polarisation. For such cases, particularly with the rapidly varying IR sky background, we should simultaneously use both ordinary and extraordinary components to measure polarisation. Simultaneous measurement of both components would make the polarisation measurement independent of the sky signal.

\section*{Acknowledgements}
We acknowledge the support provided by the observatory and technical staff at MIRO (Mt. Abu Infrared Observatory), PRL during observations. We are grateful to the night operators present at MIRO for their assistance during the observation run. We thank our colleagues in the Astronomy \& Astrophysics Division, PRL, for useful discussions and comments. We acknowledge the anonymous referee for all the valuable points which improved the quality of the paper.\\
This work has made use of data from the European Space Agency (ESA) mission
{\it Gaia} \footnote{\url{https://www.cosmos.esa.int/gaia}}, processed by the {\it Gaia}
Data Processing and Analysis Consortium (DPAC,
\footnote{\url{https://www.cosmos.esa.int/web/gaia/dpac/consortium})}.Funding for the DPAC
has been provided by national institutions, in particular the institutions
participating in the {\it Gaia} Multilateral Agreement.\\
This publication makes use of data products from the Two Micron All Sky Survey, which is a joint project of the University of Massachusetts and the Infrared Processing and Analysis Center/California Institute of Technology, funded by the National Aeronautics and Space Administration and the National Science Foundation.\\
This research has made use of the VizieR catalogue access tool, CDS, Strasbourg, France.\\
This publication makes use of data products from the Wide-field Infrared Survey Explorer, which is a joint project of the University of California, Los Angeles, and the Jet Propulsion Laboratory/California Institute of Technology, funded by the National Aeronautics and Space Administration


\bibliography{L1340.bib} 

\label{lastpage}

\end{document}